# Spatiotemporal beam stirring of femtosecond pulses in a multicore fiber driven by spectral decorrelation

NIKITA V. BOCHKAREV[1,2,†], ALENA Y. KOLESNIKOVA[1,2,†], MIKHAIL D. GERVAZIEV[1,2,*], DENIS S. KHARENKO[1,2], EVGENIY V. PODIVILOV[1,2], AND SERGEY A. BABIN[1,2,**]

[1]*Institute of Automation and Electrometry, Siberian Branch of the Russian Academy of Sciences, 1 Ac. Koptyug Avenue, Novosibirsk 630090, Russia*
[2]*Novosibirsk State University, 1 Pirogova Street, Novosibirsk 630090, Russia*
[†]*These authors contributed equally to this work.*
**gervaziev.m@gmail.com*
***babin@iae.nsk.su*

**Abstract:** Multicore fibers (MCFs) are promising for ultrafast laser systems, but random coupling between their cores makes the output power distribution unstable. Previously, we observed spatiotemporal beam stirring of narrowband sub-nanosecond pulses in a 7-core fiber, where Kerr nonlinearity equalizes the core powers. Here, we study the propagation of 150-fs pulses at 1052 nm in the same fiber. As the fiber length increases up to 20 m, the output core powers become equalized with a spread <7%, and the far-field speckle pattern transforms into a bell-shaped total beam. In contrast to the sub-nanosecond case, the effect does not depend on the input peak power (0.2–15 kW) and originates from the spectral decorrelation of the random coupling between the cores, whereas a narrowband CW source yields an unstable output with a ~40% spread. Coupled-mode simulations reproduce these results.

## 1. Introduction

Ultrafast fiber lasers are high-performance sources with the potential for robust integrated instruments used as building blocks of various photonic systems in scientific research, as well as in industrial and biomedical applications [1,2]. One of the main directions in their fundamental developments is transition to multi-channel systems which allow to overcome the limits of conventional lasers based on a singlemode fiber. For example, an ultrafast laser system with record average output power of 10.4 kW based on a coherent combination of 12 single-mode fiber amplifiers has been demonstrated in [3]. The system is able to deliver nearly transform-limited 254 fs pulses at an 80 MHz repetition rate with a high quality ($M^2 \leq 1.2$) of the combined beam. An automated spatiotemporal alignment has been developed to provide its hands-off operation. Along with the great potential of such systems, it is worth noting that their complexity increases in proportion with the number of channels.

As one of promising solutions to simplify multi-channel systems are treated to be multicore fibers (MCFs) in which multiple singlemode cores are placed in a common multimode cladding guiding a multimode pump radiation. On this way, mJ-level pulse energy and kilowatt-level average output power for a combined ultrafast laser beam after MCF amplifier has been demonstrated, see [4,5] for a review, whereas continuous-wave MCF amplifiers already offer output power beyond 1 kW [6,7]. In all the above-mentioned advanced systems, active MCFs have uncoupled cores with special measures taken for elimination any cross-talk, since even weak (random) coupling between the cores makes the output power distribution of the total beam unstable and its control becomes complicated.

Specifically, in a weakly coupled MCF, fabrication-induced differences in the core refractive indices and the perturbations caused by fiber coiling and twisting make the power exchange between the cores random, namely in specific points where core indices equalize. As a result, light launched into a single core distributes unpredictably among the cores at the output and fluctuates under external disturbances. We have recently shown that this randomness can be suppressed by Kerr nonlinearity [8]. When narrowband 435-ps pulses with a peak power of ~6 kW were launched into the central core of an 11-m-long coiled 7-core fiber (7CF), the spread of the individual core power fractions dropped from >30% to <5%. The resulting distribution became insensitive to fiber bending and twisting, and the speckled far-field pattern transformed into a stable bell-shaped total beam, reminiscent of Kerr beam self-cleaning in multimode fibers [9]. Modeling revealed the statistical nature of this spatiotemporal beam stirring effect. The nonlinear phase varying along the pulse changes the direction of energy transfer between the cores, splitting the pulse into $N \sim 100$ statistically independent sub-pulses, and averaging over them suppresses the power fluctuations by a factor of $\sqrt{N}$. The key ingredient here is the decorrelation of the core-to-core energy transfer within a single pulse, for which nonlinearity is only one possible source. For femtosecond pulses, the situation is qualitatively different: their spectrum is broad and the dispersion length is only ~0.3 m, so the pulse rapidly stretches and nonlinear phase stops accumulating after the first tens of centimeters. It thus remains an open question whether beam stirring persists for ultrashort pulses, which are of primary interest for MCF-based ultrafast laser systems [4,5,10] for various applications and especially imaging [11], and what mechanism governs it.

In this work, we address this question experimentally and numerically for ~150-fs pulses at 1052 nm propagating in the same 7CF. Using the cut-back technique for fiber lengths up to 20 m and input peak powers from 0.2 to 15 kW, we measure the power distribution among the cores in the near field, the far-field beam profile, and the temporal and spectral profiles of the pulse in the central core retrieved by frequency-resolved optical gating (FROG). We show that power equalization between the cores also occurs when using femtosecond pulses, with the spread of the core power fractions reduced to <7% at a fiber length of 20 m. In contrast to the sub-nanosecond case, the effect is independent of the input peak power and originates from the spectral decorrelation of the linear random coupling between the cores. The experimental results are reproduced by a coupled-mode model that accounts for group-velocity dispersion and the refractive index perturbations caused by fiber coiling and twisting.

## 2. Experiment

In the experiments, the radiation source was an AVESTA TEMA-100 ytterbium solid-state femtosecond laser. It operates at a central wavelength of 1052 nm and delivers ~150 femtosecond pulses at a repetition rate of 80 MHz. The laser beam was coupled into a passive 7CF fabricated at Dianov Fiber Optics Research Center, Prokhorov General Physics Institute of the Russian Academy of Sciences (see [8] for details). At 1052 nm, each core of the 7CF behaves as a single-mode fiber with a mode field diameter of 6.4 μm, and the core-to-core spacing is 17.3 μm. The experimental setup is shown schematically in Fig. 1. To maximize coupling efficiency, the waist diameter of the focused beam was matched to the mode field diameter of a single core. Under optimal focusing into a core, the transmission of the system reached 55%. The distribution of the output power among the cores was measured with a Gentec Beamage-4M beam profiler. The temporal profile and spectrum of the pulse were measured using a Femto Easy MS-FROG-SLP (FROG). For this purpose, an iris diaphragm placed in front of the FROG selected the central beam out of the seven beams in the near-field output profile, so that only the pulse from the central core was measured.

To study the spatiotemporal and spectral dynamics of light during its propagation in a 7CF, we performed an investigation of the beam evolution along the fiber via gradual back-side cutting under unchanged launch conditions. To average the influence of refractive index fluctuations along a core on the studied spatial dynamics, a fiber piece with an initial length of 20 m was coiled into a ring with a diameter of 6 cm. The output power distribution across the fiber was examined both in continuous-wave (CW) laser operation ($P_{av}$=40 mW) and in femtosecond pulse mode, with input peak power $P_{in}$ values ranging from 0.2 kW to 15 kW, taking transmission into account.

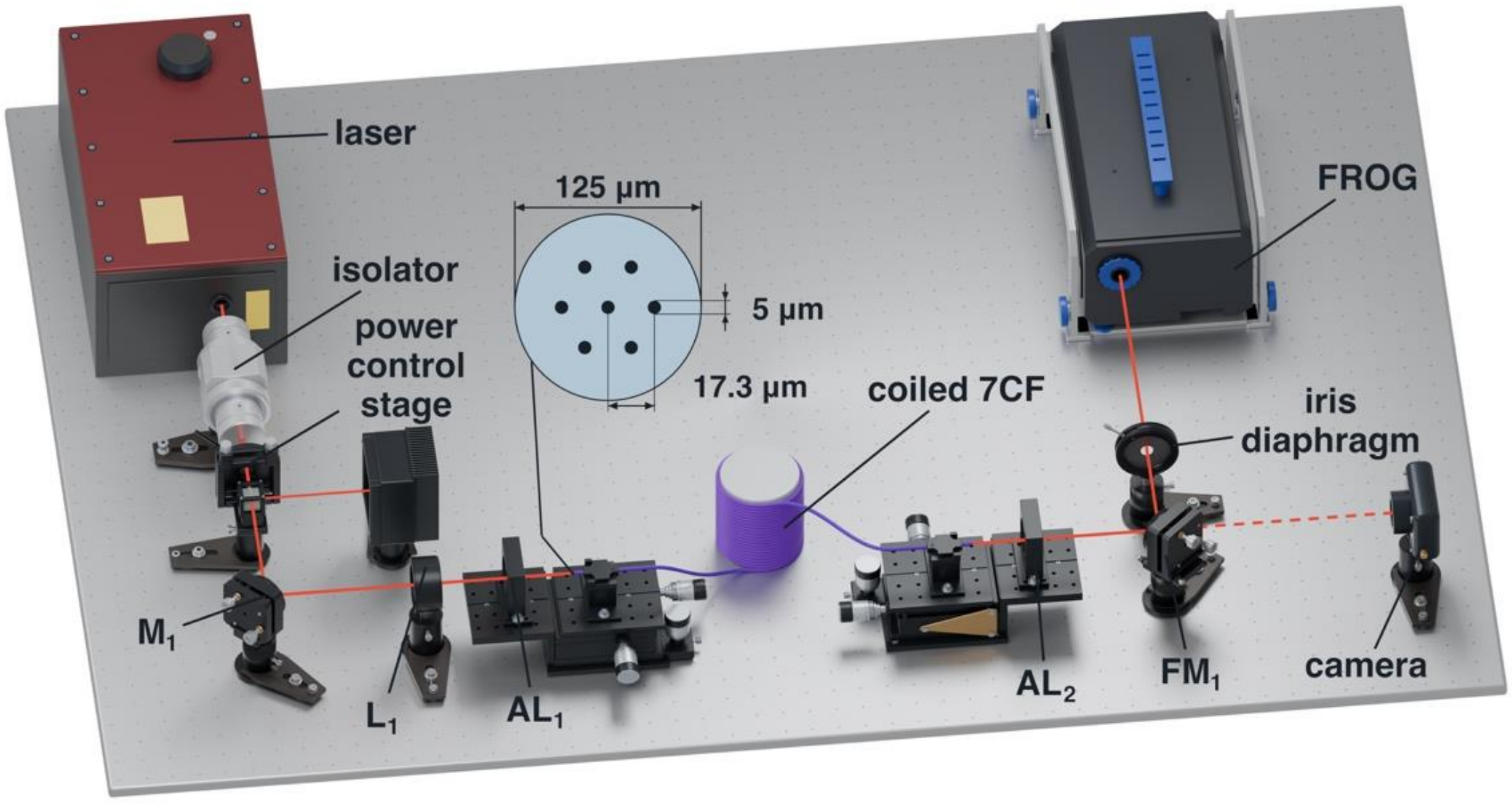

Fig. 1. Experimental setup: femtosecond laser, optical isolator, power control stage (half-wave plate, polarizing beam splitter, and beam dump), spherical lens $L_1$, input aspheric lens $AL_1$, coiled seven-core fiber (7CF), output aspheric lens $AL_2$, flip mirror $FM_1$, iris diaphragm, FROG and CMOS camera.

The obtained experimental data were used to plot the difference between the maximum and minimum output power fraction in the cores as a function of the input peak power (Fig. 2a) similar to experiments with narrowband sub-ns pulses [8]. The graph presents results for several fiber lengths. The first column of data points corresponds to the case of CW laser operation. For a fiber length of 2 m, a noticeable dependence on the input power is observed, whereas for fiber lengths starting from 5 m and above, the difference between the maximum and minimum output power becomes independent of the input power, indicating the linear nature of the observed effect. At the same time, with increasing fiber length, a decrease in the measured spread of the output power fractions is observed, down to <7% for a fiber length of 20 m. Thus, the spatial dynamics depend significantly on the fiber length. Figure 2b presents the relative power difference in the cores starting from the input face (corresponding to the pulse launch into the central core) up to 20 m in length for three different levels of $P_{in}$. For the input peak power levels of 0.2 kW, 5 kW and 13 kW, a qualitatively similar spatial dynamics is observed, in which a clear reduction in the output power spread with increasing fiber length is evident.

We also studied the evolution of the transverse beam profile when launching pulsed radiation with a peak power of 13 kW into fibers of different lengths. Near-field (NF) and, using a Fourier lens, far-field (FF) images were captured (Fig. 3). For the 2-m-long fiber (Fig. 3(a)), the NF power distribution among the cores is random and unstable, with the maximum power concentrated in one core, while some other cores contain almost no power. For the 20-m-long fiber (Fig. 3c), stable equipartition was established, and all cores exhibit nearly equal output power. At the same time, the FF images demonstrate noticeable evolution during the transition from the 2-m fiber length to the 20-m length. It is noteworthy that for the 2-m fiber length (Fig. 3d), the beam exhibits a speckled structure, whereas for the 20-m length the beam acquires a characteristic bell-shaped profile (Fig. 3f).

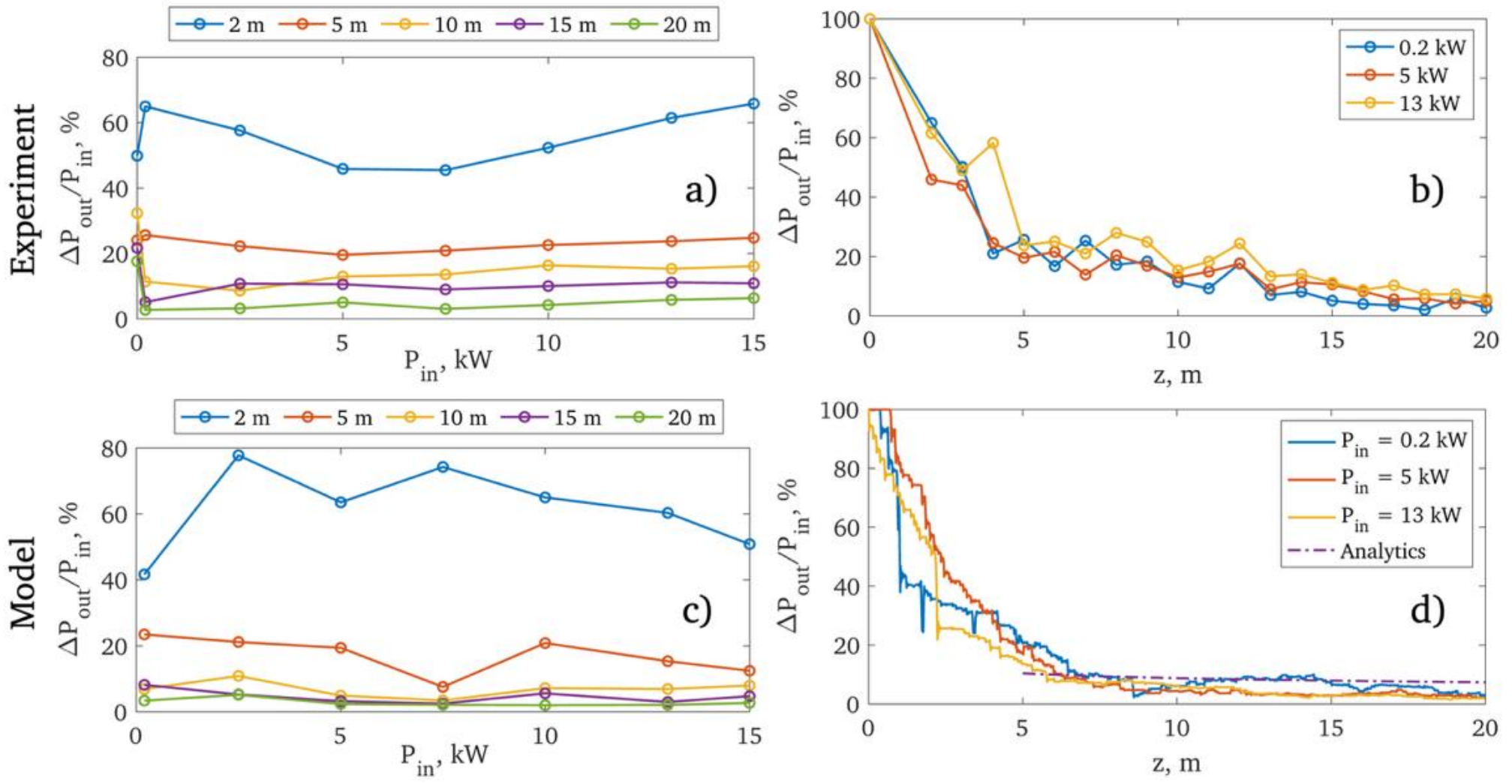

Fig. 2. Relative difference between the maximum and minimum output power among the cores as a function of the input peak power in experiment (a) and modeling (c). Relative difference between the maximum and minimum output power as a function of the fiber length in experiment (b) and modeling (d) for three input peak power values: 0.2 kW (blue), 5 kW (red), 13 kW (yellow).

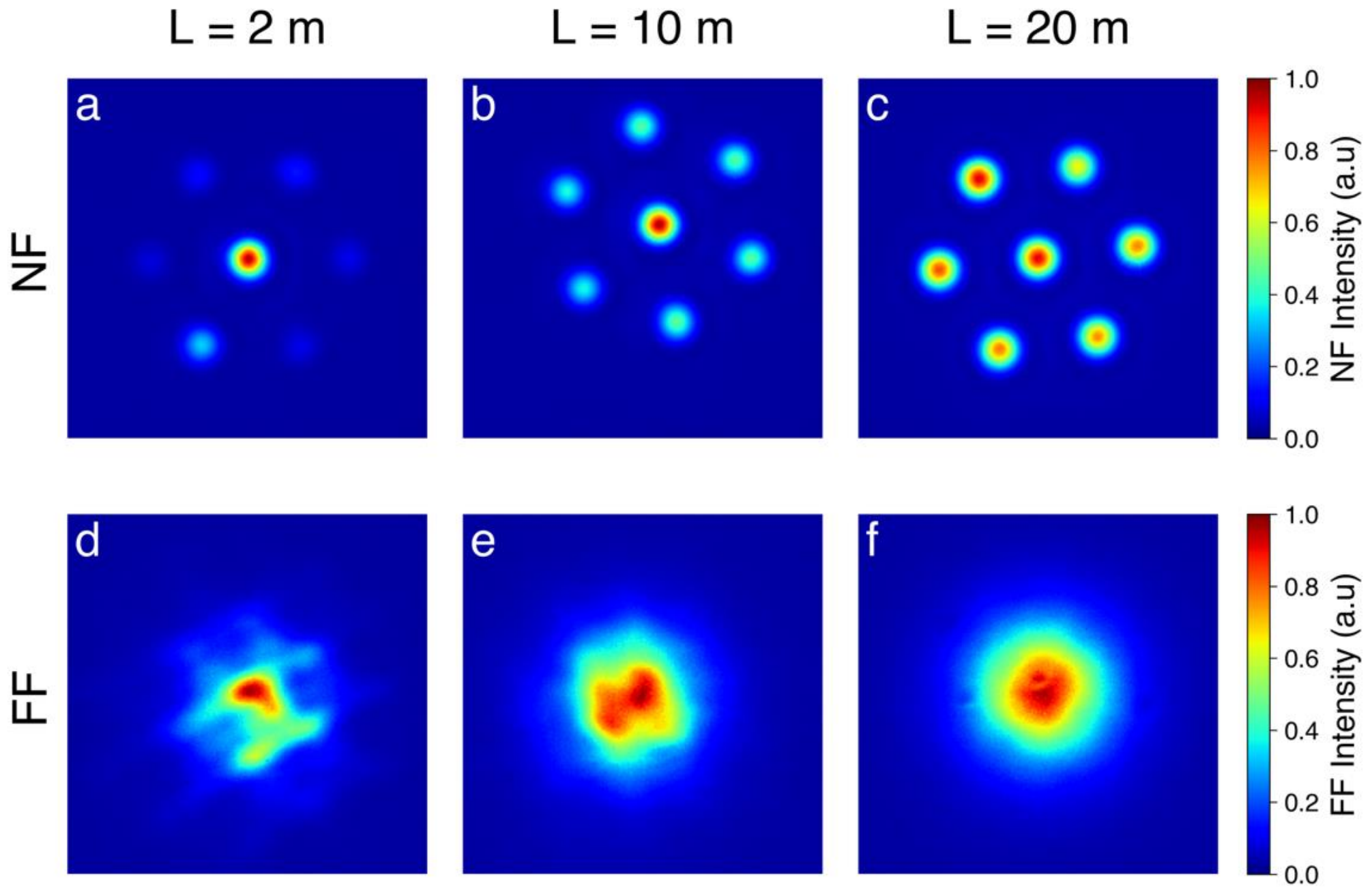

Fig. 3. Near-field (NF) images of the 7CF output captured at 1052 nm (a–c) demonstrate the formation of 7 beams with equal and stable intensities when the fiber length increases. Far-field (FF) images (d–f) demonstrate the transition from speckled structure to the bell-shaped transverse profile of the total beam. Input peak power $P_{in}$ = 13 kW.

Finally, to study the nature of this effect in detail, the temporal and spectral profiles of the central beam were retrieved from the measured FROG traces (Fig. 4a,c) using a standard iterative phase-retrieval algorithm. The retrieved temporal and spectral profiles are presented in Fig. 4e,f. The input pulse has a duration of 157 fs (FWHM) and a spectral width of 10.7 nm.

The corresponding time–bandwidth product of 0.46 exceeds the transform limit of a sech² pulse (0.315) by a factor of ~1.4, indicating a small residual chirp. After only 2 m of fiber, the pulse broadens to about 2.6 ps and acquires a pronounced multi-peak structure (Fig. 4e). At the same time, its spectrum roughly doubles in width, to ≈ 23 nm, and becomes strongly modulated (Fig. 4f). With further propagation, the pulse continues to stretch almost linearly with fiber length, at about 2 ps/m. It reaches ≈6 ps at 5 m, ≈16 ps at 10 m, ≈24 ps at 15 m, and ≈ 40 ps at 20 m. Over the same lengths, the spectral width remains within 23–29 nm interval, and the spectral centroid shifts only slightly, from 1051 to 1054 nm. This suggests that spectral broadening occurs mainly within the first few tens of centimeters of fiber, where the peak power is still high. As the pulse disperses, its peak power drops, the nonlinear interaction weakens, and the spectrum stops evolving. The temporal profile, however, continues to broaden becoming strongly chirped due to dispersion, and fragments into a dense train of sub-picosecond peaks.

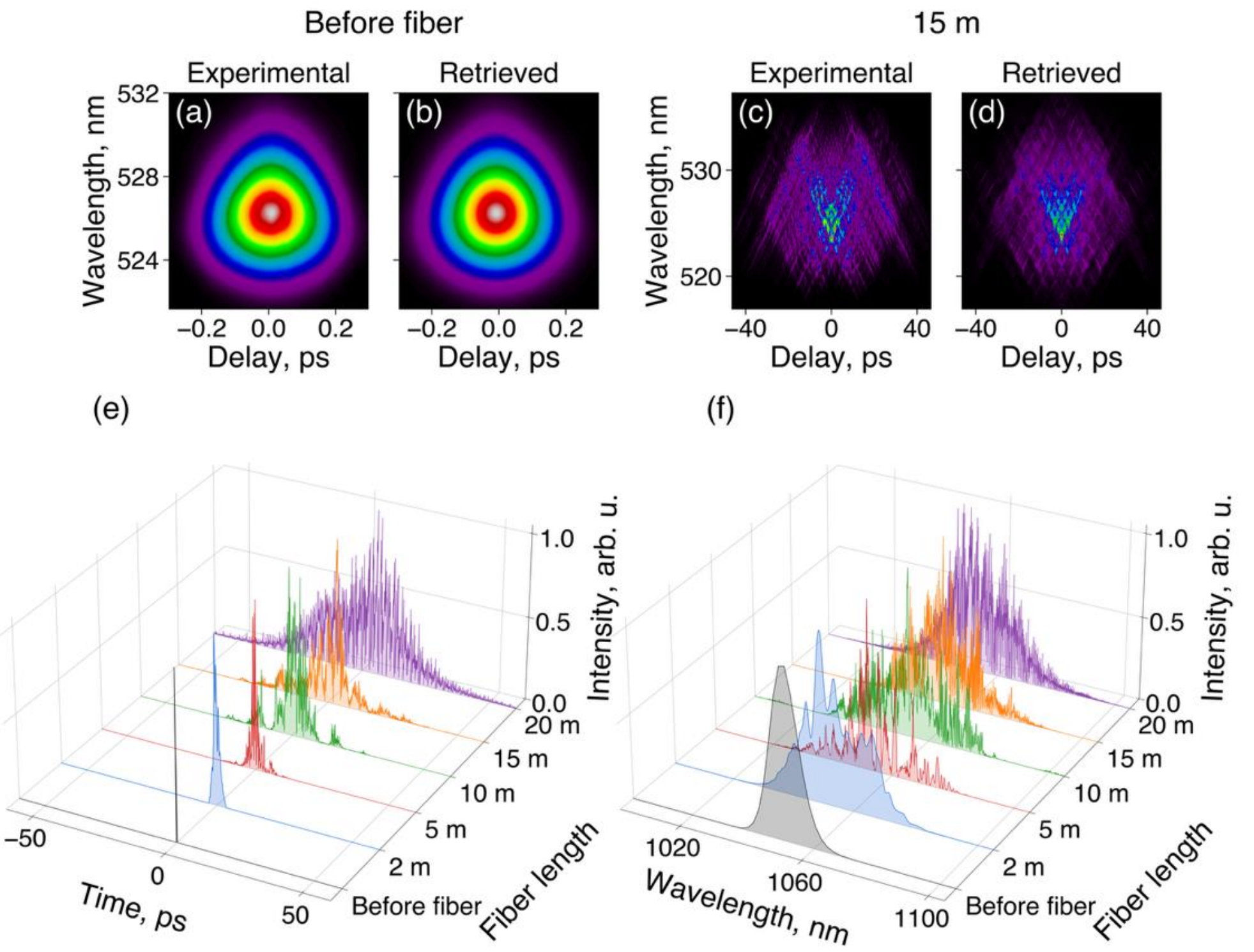

Fig. 4. Experimental and retrieved SHG-FROG traces: (a), (b) before the fiber; (c), (d) after 15 m. Evolution of (e) the temporal intensity profile and (f) the optical spectrum of the pulse with increasing fiber length (0, 2, 5, 10, 15 and 20 m) at an input peak power of 13 kW.

## 3. Model

The pulse propagation model in a multicore fiber can be written as follows. The electric field is the sum of the pulses propagating in the cores $E = \sum_{j=0}^{6} A_j(z,t)\Psi(r - r_j)e^{i\omega t - ikz}$ with the carrier frequency $\omega = 285$ THz ($\lambda = 1052$ nm) $k = \frac{2\pi n_{\text{eff}}}{\lambda} = 8.6\ \mu\text{m}^{-1}$, where $n_{eff}$ is the average refractive index in cores. Here $\Psi(r - r_j)$is the normalized spatial distribution of core j fundamental mode in fiber cross-section, $r_j$ is the coordinate of core j center (j=0,...,6 with j=0 for central core). The strength of the coupling between the modes of different cores is calculated from perturbation theory ((6) in [12]) and for our experimental geometry of MCF is estimated as $J = 1.15 \cdot 10^{-6}$. Propagation evolution of field amplitudes is described by a system of coupled equations:

$$\begin{cases} \frac{dA_0}{dz} = i\left(k + \frac{in_{\mathrm{eff}}}{c}\frac{d}{dt}\right)\left[J\sum_{j=1}^{6} A_j + \delta n_0(z) A_0\right] + i\gamma |A_0|^2 A_0 + i\frac{\beta_2}{2}\frac{\partial^2 A_0}{\partial t^2}, \\ \frac{dA_j}{dz} = i\left(k + \frac{in_{\mathrm{eff}}}{c}\frac{d}{dt}\right)\left[J\left(A_0 + A_{j-1} + A_{j+1}\right) + \left(\delta n_{\mathrm{j}}^{\mathrm{tw}}(z) + \delta n_j(z)\right) A_j\right] + i\gamma \left|A_j\right|^2 A_j + i\frac{\beta_2}{2}\frac{\partial^2 A_j}{\partial t^2}. \end{cases} \quad (1)$$

Here the square of the field amplitude is normalized to the power. The coefficient of self-phase modulation due to Kerr nonlinearity is $\gamma = 5 \cdot 10^{-3} (\mathrm{W} \cdot \mathrm{m})^{-1}$ in each core, $\beta_2 = 24 \frac{\mathrm{ps}^2}{\mathrm{km}}$ is the group velocity dispersion parameter. The addition of a time derivative is necessary to account for the spectral dependence of phase accumulated in different cores i.e., the inter-core group delay. The model includes the relative difference in refractive indices $\delta n_j(z)$ in the cores arising at fiber fabrication (it is randomly varied such that $\sum_{j=0}^{6} \delta n_j(z) = 0$ and does not exceed a certain specified value $\left|\delta n_j\right| < \delta n_{\mathrm{max}} = 1.5 \cdot 10^{-6}$) and relative difference in refractive index caused by coiling fiber into a ring $\delta n_j^{\mathrm{tw}}(z) = \frac{0.78\rho}{R}\cos\left(\frac{\pi j}{3} + \varphi_j(z)\right)$ [13], here coefficient 0.78 is caused by photoelastic effect [14], $R = 3$ cm is the radius of the ring, $\rho = 17.3$ μm is the distance between the cores, $\varphi_j(z)$ is the random angle of twist of the fiber around its axis. Important note: the change in the refractive index caused by coiling fiber $\delta n_j^{\mathrm{tw}} < 4.5 \cdot 10^{-4}$ is two orders of magnitude larger than $\delta n_{\mathrm{max}}$ and has a dominant effect. The initial state is the pulse injected into the central core at the fiber input which corresponds to the experimental input pulse.

The numerical scheme for solving the system of equations (1) is based on the Split-Step method, where the linear and nonlinear parts of the equation are calculated independently; the linear part is calculated using the matrix exponent. To calculate the time derivative in a step with a linear coupling of cores, the Fourier transform in time was used. The time step is $\tau = 150$ fs, the spatial step is $h = 250$ μm. The relative difference in refractive indices $\delta n_j(z)$ and $\varphi_j(z)$ random change with a characteristic length $L_{\mathrm{dn}} = 1$ mm and $L_{\varphi} = 30$ cm, the intermediate values $\delta n_j(z)$ and $\varphi_j(z)$ are determined by a spline.

As in the experiment, we ran simulations for several input peak powers and observed the same trend in the output intensity distribution among the cores (Fig. 2c,d). Because the rate at which the intensity equalizes across the cores does not depend on the input peak power both in theory and experiment (Figs. 2b,d), it is confirmed that the observed effect is mainly linear. Nonlinear spectral broadening occurs only within the first few tens of centimeters. After that, the pulse broadens in time and becomes chirped as it propagates through the normally dispersive fiber, so its peak power drops and nonlinear effects become negligible.

It is important to note that the nonlinear broadening predicted by the modeling exceeds the experimental results by approximately a factor of 1.5 (Fig. 5b1-2, d1-2). We attribute this discrepancy to the effect of spectral filtering—with a characteristic width on the order of 40 nm—that arises from signal processing via FROG caused by the finite phase-matching bandwidth of the second-harmonic-generation crystal used in the FROG setup [15]. We applied a filter with a Lorentzian-shaped window to the model spectra and obtained a reasonably good agreement of the filtered model spectra (Fig. 5f1-2) and corresponding temporal profiles (Fig.5e1-2) with the experimental data (Fig. 5 a1-2, b1-2).

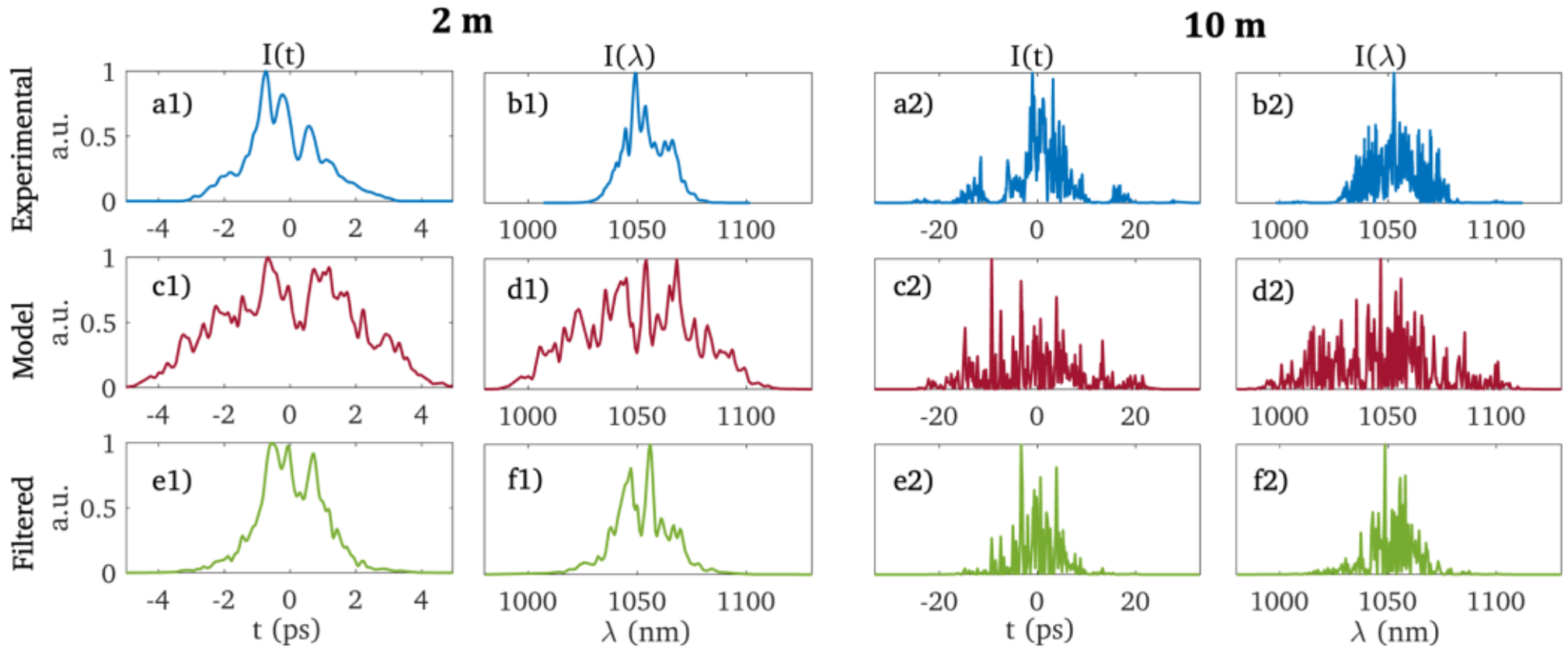

Fig. 5. Comparison of the central-core temporal intensity profiles in experiment (a1-2), model (c1-2) and spectral-filtered model (e1-2) and corresponding optical spectra of the pulse in experiment (b1-2), model (d1-2) and spectral-filtered model (e1-2) for fiber length of 2 m (1) and 10 m (2) at the input peak power of 13 kW.

## 4. Discussion

So, the primary cause of beam stirring between the cores - and the resulting equalization of intensity among them for the present experiment with fs pulses - is independent stochastic mixing of different spectral components, which becomes noticeable for a femtosecond pulse due to its large spectral width, in contrast to the narrowband sub-ns pulses [8]. Random perturbations along the fiber, such as bends whose direction changes randomly every few tens of centimeters, add a random index offset $\delta n_{\mathrm{j}}^{\mathrm{tw}}(z)$ to each core. These offsets constantly detune the cores from one another, so instead of coherent, periodic power exchange, light hops between the cores randomly in specific points where core indices equalize. After a mixing length $L_{\mathrm{eq}}$ the light has "forgotten" which core it started in. All propagation phases, including those set by the random index offsets and by the coupling, scale with frequency ($\varphi \sim k - \frac{n_{\mathrm{eff}}}{c}\Delta\omega$). Different spectral components distributed along a highly-chirped pulse therefore travel through the same random sequence of perturbations but accumulate slightly different phases, and once these differences reach about 1 radian, the components are mixed between the cores independently of each other. In the time domain, the same physics appears as a spread of group delays $\Delta\tau_c$ between the different paths the light can take. The two pictures are linked: due to the large chirp two decorrelated sub-pulses are separated by group delay $\Delta\tau_c = \beta_2 L \Omega_{\mathrm{c}}$ .

The power fraction of the whole pulse in a given core is the average of the per-frequency fractions distributed over its spectrum and length respectively, since the pulse is chirped. If the spectral width $\Delta\omega$ spans many decorrelation bandwidths, the pulse effectively contains $N \sim 1 + \frac{\Delta\omega}{\Omega_{\mathrm{c}}}$ independent sub-spectra, and equivalently $N$ sub-pulses in time domain. Their random fluctuations partially cancel, so the pulse-energy fraction fluctuates much less than that of a single frequency: $\frac{\Delta P(z)}{P_{\mathrm{in}}} = \frac{\Delta P(0)}{P_{\mathrm{in}}}\frac{1}{\sqrt{N}}$.

Let us estimate the spectral decorrelation bandwidth $\Omega_{\mathrm{c}}$. For cores $i$ and $j$ at frequency $\omega_0 + \Delta\omega$ , the phase shift is: $\Phi_{ij}(\omega_0 + \Delta\omega) \approx \Phi_{ij}(\omega_0) + \Delta\omega X_{ij}$ , where $X_{ij} = \frac{n}{c}\int_0^z [\delta n_{\mathrm{tw}}^i(z') - \delta n_{\mathrm{tw}}^j(z')]dz'$. Two frequencies separated by $\Delta\omega$ differ in phase by $\Delta\omega X_{ij}$. When this shift is on the order of 1 radian, the interference of paths between the cores is rearranged, and the frequencies are distributed among the cores independently. This defines the decorrelation width $\Omega_{\mathrm{c}} = \frac{1}{\sqrt{\langle X_{ij}^2 \rangle}}$. Assuming the phase change $\varphi_j(z)$ to be a random walk with a characteristic length $L_\varphi$, we can estimate $\sqrt{\langle X_{ij}^2\rangle} \sim \frac{0.78\rho n}{Rc}\sqrt{zL_\varphi}$. Thus, the power spread between the cores decreases with the pulse propagation distance along the fiber as follows:

$$\frac{\Delta P(z)}{P_{\text{in}}} = \frac{\Delta P(0)}{P_{\text{in}}}\left(1 + \frac{0.78\rho n}{Rc}\sqrt{zL_\varphi}\Delta\omega\right)^{-1/2} \quad (2)$$

This analytical approach is in good agreement with the experimental and numerical data for the dependence on the propagation distance (Fig.2d) taking into account that it is applicable for relatively long distances.

So, when a pulse is launched as a whole into randomly coupled cores, it is split among them randomly, and the power in each core can differ from the mean value by about the mean value itself. If, however, the pulse can be divided into *N* parts which are split among the cores independently, these random differences partly cancel out. They become $\sqrt{N}$ times smaller after averaging over the whole pulse, and each core receives almost exactly 1/7 of the power even in a single-pulse measurement. Two parts are split independently if the phase mismatch between the cores differ for them by more than ~1 radian. In the linear effect, the phase changes across the spectrum: the index difference $\delta n_{\text{tw}}$ is the same for the whole pulse, but the phase it adds is proportional to frequency changing linearly along the chirped pulse. Here the parts occupy spectral intervals of width $\Omega_c$, so a pulse with broad spectrum will have large number of statistically independent sub-pulses *N*, while the power does not matter.

In the nonlinear stirring effect reported in [8], the spectrum of sub-nanosecond pulse is narrow, so the linear effect does not work there. However, due to the Kerr effect, each core gains a nonlinear phase proportional to its own instantaneous power. This phase is different in different cores and at different points of the pulse, so the pulse itself creates the index difference between the cores, and it is different for each part of the pulse. Here the parts are time intervals with different power, and their number $N \approx \gamma P_{in}\, z/7$ grows linearly with propagation length and with peak power value. This explains the practical differences between the two kinds of beam stirring effect. The linear effect needs a broad spectrum but does not depend on power. The nonlinear effect works with narrowband pulses but needs high peak power, and since its randomness comes from the pulse itself, it is not sensitive to external disturbances of the fiber.

## 5. Conclusion

In conclusion, we have studied the propagation of ~150-fs pulses centered at 1052 nm in a seven-core fiber coiled into a ring. When the pulse is launched into the central core, the spread of the output core power fractions decreases with fiber length to <7% at 20 m, and the far-field pattern transforms from a speckled structure into a stable bell-shaped total beam, whereas narrowband CW radiation propagating in the same fiber is measured to have an unstable output with a ~40% spread.

In contrast to the narrowband sub-nanosecond case [8], the observed equalization does not depend on the input peak power in the range of 0.2–15 kW. FROG measurements show that the pulse in the central core stretches to ~40 ps with increasing fiber length to 20 m and fragments into a dense train of sub-picosecond peaks, whereas its spectrum broadens to 23–29 nm within the first meters only. The coupled-mode model that accounts for dispersion, the inter-core group delay, and the random coiling- and twist-induced index perturbations reproduces both the spatial and the temporal dynamics.

Together with our previous results [8], these findings show that the beam stirring effect in MCFs is a general statistical phenomenon that occurs whenever the core-to-core energy transfer becomes decorrelated within a single pulse. The pulse then effectively splits into *N* independent parts, each distributed among the cores randomly, and averaging over the whole pulse suppresses the power fluctuations among the cores by a factor of $\sqrt{N}$. The two variants of the effect differ in what creates these independent parts within the pulse. In the linear (femtosecond) variant, the parts are spectral intervals of width $\Omega_c$, which is set by the coiling-induced inter-core group delay. Hence, their number *N* within the chirped pulse grows with the spectral width and the fiber length [Eq. (2)] but not with power, and a broad spectrum is required. In the nonlinear (sub-nanosecond) variant, the parts are time intervals created by the self-induced Kerr phase proportional to instantaneous power, their number is estimated as $N \approx \gamma P_{in} z/7$, so the effect works for narrowband pulses but requires a high peak power.

In addition, it is worth discussing that the beam stirring shares its main observable signature with Kerr beam self-cleaning effect in graded-index (GRIN) multimode fibers for sub-nanosecond pulses [9] and for femtosecond pulses as well [16], where a speckled output beam also transforms into a stable bell-shaped one in both cases. However, the underlying physics, despite some similarities, may have important differences. Beam self-cleaning is an intrinsically nonlinear threshold effect in both cases, in which four-wave mixing irreversibly redistributes energy among the transverse modes of GRIN fiber toward a Rayleigh–Jeans (RJ) distribution dominated by the fundamental mode, as described by optical wave thermodynamics [17,18], so that the spatial coherence of the output beam increases. In beam stirring, energy is not transferred to any preferred spatial mode of an MCF. Instead, the pulse energy is equally distributed among the cores, and the fields in different cores become decorrelated for sub-pulses formed within single pulse. At the same time, averaging over statistically independent sub-pulses may justify the applicability of thermodynamic approach [8]. In terms of thermodynamic description, the equipartition induced by spatio-temporal beam stirring in nonlinear case may be also associated with the RJ distribution at a temperature tending to infinity, but in MCF it is applicable to supermodes with different propagation constants as discussed in [8, 19], whereas in our case of weakly coupled cores it is more relevant to treat degenerate core modes for which equipartition may be reached at any temperature. It should be noted that the equipartition for MCF modes observed in linear case qualitatively corresponds to the equipartition of transverse modes in multimode GRIN fiber observed for broadband multimode laser diode radiation which transforms with increasing length to beam with a parabolic intensity profile in a broad power range, see e.g. [20].

Thus, the revealed beam stirring effect in multicore fibers behaves similarly to the beam self-cleaning effect in graded-index multimode fibers but has important distinctive features. The performed detailed analysis of the experimental data supported by numerical simulations, reveals fundamentally different mechanisms for the long narrowband and short broadband nearly transform limited pulses. In these two cases dominate either nonlinear or linear effects, respectively, both leading to variation of phase incursion within single pulse and corresponding splitting on sub-pulses randomly distributed among the cores. As a result of statistical averaging for the whole pulse, the power distribution among the cores becomes equalized and stabilized in both cases. In intermediate cases of picosecond-scale pulses, both effects may interplay together thus offering additional opportunities to control pulse propagation/distribution in a MCF. The studied effect may be important in MCF amplifiers [4-7], CW and pulsed MCF lasers with coupled cores [21-25], as well as for MCF applications in material processing, sensing, communication, biomedical diagnostics and imaging [1,2,10,11], as well as other laser technologies.

**Funding.** Russian Science Foundation (21-72-30024-П); Ministry of Science and Higher Education of the Russian Federation (FSUS-2025-0010).

**Disclosures.** The authors declare no conflicts of interest.

**Data availability.** Data underlying the results presented in this paper are not publicly available at this time but may be obtained from the authors upon reasonable request.